\def\bq{\begin{equation}}
\def\eq{\end{equation}}
\def\bqa{\begin{eqnarray}}
\def\eqa{\end{eqnarray}}
\def\bqb{\begin{eqnarray*}}
\def\eqb{\end{eqnarray*}}
\documentstyle[12pt]{article}\textwidth 16cm
\def\pr#1#2#3{ Phys. Rev. ${\bf{#1}}$ (#2) #3}

\def\pl#1#2#3{ Phys. Lett. ${\bf{#1}}$ (#2) #3 }

\def\np#1#2#3{ Nucl. Phys. ${\bf{#1}}$ (#2) #3}

\global\nulldelimiterspace = 0pt

\def\Bsl{\hbox{/\kern-.6700em$B$}} 
\def\Dsl{\hbox{/\kern-.6700em$D$}} 
\def\Wsl{\hbox{/\kern-.6700em$W$}} 

\def\roughly#1{\mathrel{\raise.3ex
    \hbox{$#1$\kern-.75em\lower1ex\hbox{$\sim$}}}}

\def\gsim{\roughly>}

\def\mh2{m^2_H}

\begin{document}
\pagenumbering{arabic}
\thispagestyle{empty}
\hspace {-0.8cm} hep-ph/9609520\\
\hspace {-0.8cm} PM/96-27\\
\hspace {-0.8cm} September 1996\\
\vspace {0.8cm}\\

\begin{center}
{\Large\bf     Longitudinal Polarization at future $e^+e^-$ Colliders}\\
\vspace {0.1cm}
{\Large\bf  and Virtual New Physics Effects} \\

 \vspace{1.8cm}
{\large F.M. Renard$^a$ and C.
Verzegnassi$^b$}
\vspace {1cm}  \\

$^a$Physique
Math\'{e}matique et Th\'{e}orique,
UPRES-A 5032,\\
Universit\'{e} de Montpellier II,
 F-34095 Montpellier Cedex 5.\\
\vspace{0.2cm}
$^b$ Dipartimento di Fisica,
Universit\`{a} di Lecce \\
CP193 Via Arnesano, I-73100 Lecce, \\
and INFN, Sezione di Lecce, Italy.\\

\vspace{1.5cm}

 {\bf Abstract}
\end{center}
\noindent

The theoretical merits of longitudinal polarization asymmetries of
electron-positron annihilation into two final fermions at future
colliders are examined, 
using a recently proposed theoretical description. A number of
interesting features, valid for searches of virtual effects of new
physics, is underlined, that is reminiscent of analogous
properties valid on top of $Z$ resonance. 
As an application to a concrete example, we consider the case of
a model with triple anomalous gauge couplings and show that the
additional information provided by these asymmetries would lead to a
drastic reduction of the allowed domain of the relevant parameters.

\vspace{1cm}

\setcounter{page}{0}
\def\thefootnote{\arabic{footnote}}
\setcounter{footnote}{0}
\clearpage

\section{Introduction}\par

Among the several quantities that can be measured in the process of
electron-positron annihilation into a fermion-antifermion couple, the
longitudinal polarization asymmetry $A_{LR} \equiv
{\sigma_{L}-\sigma_{R}\over \sigma_{L}+\sigma_{R}}$ has
represented in the last few years an example of, least to say,
remarkable theoretical interest. This is due to the known fact that, as
it was stressed in a number of dedicated papers \cite{1},\cite{2},
\cite{3},\cite{4}, the properties of this observable on top of Z
resonance are indeed special. In particular one can
stress two main facts i.e. that $A_{LR}$ is independent of the final
produced state (this was shown in particular detail in Ref.\cite{3}),
and that it is particularly sensitive to possible virtual effects of a
large number of models of new physics (this was exhaustively discussed
in Refs.\cite{2} and \cite{4}). These features, that appear essentially
unique, have deeply motivated the tough experimental effort at SLC
\cite{5} where $A_{LR}$ has been (in fact, it is still being)
measured to an extremely high precision \cite{6}, fully exploiting the
fact that at a linear electron-positron collider it is "relatively"
easy to produce longitudinally polarized electron beams with a high and
accurately known polarization degree\cite{7}. This is not the case of a
circular accelerator, and for this reason neither at LEP1 (in spite of
the several impressive experimental studies and efforts of recent years
\cite{8}) nor at LEP2 a measurement of $A_{LR}$ has been, or will be
predicticably performed.\par
The possibility that a linear electron-positron collider of an overall
c.m. energy not far from 500 GeV is eventually built in a not too
distant future has been very seriously investigated in the last few
years, and the results of a remarkable combined experimental and 
theoretical effort have been published in several dedicated
Proceedings\cite{8bis}. At such a kind of machine it would be, again,
"relatively" easy to produce longitudinally polarized electron beams,
which implies the possibility of measuring $A_{LR}$, for
various possible final states. One might therefore wonder whether the
special theoretical properties valid on top of Z resonance will still
be true and, if not, how they would be modified at about 500 GeV.\par

The purpose of this paper is precisely that of investigating the
general features of $A_{LR}$ at such a future linear collider (NLC) and
to show that, from a theoretical point of view, this quantity still
retains beautiful and interesting features, that make it particularly
promising as a tool for investigating virtual effects of models of new
physics. This will be shown in some detail in the following
Section 2. Section 3 will be devoted to an illustrative example, i.e.
to the case of a model with anomalous gauge couplings, for
which the benefits of a measurement of $A_{LR}$ will be explicitely
shown in a quantitative way. A short  final discussion will then be
given in Section 4, valid for a more general class of theoretical models.

\newpage                     

\section{Longitudinal polarization asymmetries at one loop.}

\vspace{0.25cm}

{\bf 2a. General features}

\vspace{0.25cm}

The purpose of this section is that of deriving relatively simple and
compact expressions for longitudinal polarization asymmetries in the
process of electron-positron annihilation into two final fermions at
arbitrary c.m. energy at one loop. With this aim, we shall follow a
procedure that has been fully illustrated in two recent papers
\cite{9},\cite{10} and has been called "$Z$-peak subtracted"
representation. In order, though, to make this paper, at least
reasonably, self-contained, we shall sketch a quick derivation of all
the fundamental formulae, defering to refs.\cite{9},\cite{10} for a
more complete discussion of several technical details.\par

The starting point of our derivation will be the expression of the
longitudinally polarized cross sections $\sigma_{L,f}$ 
and $\sigma_{R,f}$
(left-handed and right-handed initial electrons) in Born approximation,
where $f$ denotes the final fermion (in the case that we shall
consider, charged lepton or quark). In practice, though, it will be
more useful to consider from the very beginning the difference and the
sum of such cross sections, that appear directly as the numerator and
the denominator of the longitudinal polarization asymmetry. Denoting by
$\sigma_{LR,f}$ and $\sigma_f$ these quantities, one easily finds that

\bqa &&
\sigma^{(0)}_{LRf}(q^2)=\sigma^{(0)}_{Lf}(q^2)-\sigma^{(0)}_{Rf}(q^2)
=N_f({4\pi q^2\over3})\times
 \nonumber\\
&&\{[{\sqrt2 G^{(0)}_{\mu}
M^2_{Z0}\over4\pi}]^2{2g^{(0)}_{Vl}g^{(0}_{Al})
(g^{(0)\, 2}_{Vf}+g^{(0)\, 2}_{Af})\over(q^2-M^2_{Z0})^2}-
2{\alpha_0 
Q_f\over q^2}[{\sqrt2 G^{(0)}_{\mu}
M^2_{Z0}\over4\pi}]
{g^{(0)}_{Al}g^{(0)}_{Vf}\over q^2-M^2_{Z0}}\} \eqa

\bqa &&
\sigma^{(0)}_{f}(q^2)=\sigma^{(0)}_{Lf}(q^2)+\sigma^{(0)}_{Rf}(q^2)
=N_f({4\pi q^2\over3})\{\{{\alpha^2_0  Q^2_f\over q^4}+ \nonumber\\
&&
[{\sqrt2 G^{(0)}_{\mu}
M^2_{Z0}\over4\pi}]^2{(g^{(0)\, 2}_{Vl}+g^{(0\, 2}_{Al})
(g^{(0)\, 2}_{Vf}+g^{(0)\, 2}_{Af})\over(q^2-M^2_{Z0})^2}- 
2{\alpha_0 
Q_f\over q^2}[{\sqrt2 G^{(0)}_{\mu}
M^2_{Z0}\over4\pi}]
{g^{(0)}_{Vl}g^{(0)}_{Vf}\over q^2-M^2_{Z0}}\} \eqa

In the previous formulae, $q^2$ is the total c.m. squared energy, 
$N_f$ is the
colour factor and
the various couplings are defined in the conventional way,
i.e  $g^{(0)}_{Al,f} \equiv I^{3L}_{l,f}$ ;  
$g^{(0)}_{Vl,f}\equiv  I^{3L}_{l,f} -2Q_{l,f}s^{2}_{0}$, with $Q_{l,f}$
the charge of the lepton $l$ or fermion $f$.
Note that all the couplings and the $Z$ mass (with index (0)) are, 
by definition, 'bare'
ones.\par

From eqs.(1),(2) it is straightforward to derive the expression at Born
level of the longitudinal polarization asymmetry $A^{(0)}_{LR,f}(q^2)$
defined as $\sigma^{(0)}_{LR,f}/\sigma_f$.

From a glance to eqs.(1),(2) one can derive a rather important and
well-known fact. On top of $Z$ resonance, 
where the pure $Z$ exchange term
largely dominates, the dependence on the final state completely
disappears, so that $A^{(0)}_{LR,f}$ becomes 
only dependent on the initial
electron-$Z$ couplings. But when one moves away from the $Z$ peak this
peculiar feature disappears, and other terms become competitive. As a
result of this, $A^{(0)}_{LR,f}$ will now 
effectively depend on products of
$Z$ couplings to the initial and to the final
considered fermion, and several different observables will therefore
become potentially relevant.\par
Concerning the final fermion, we shall be limited in this paper to the
case of "light" charged ones ($f=l,u,d,s,c,b$). Moreover, the considered
$q^2$ values will always be (much) larger than $M^2_Z$. In terms of the
final masses, this means that they will be safely negligible,
$m_f\simeq 0$. For what concerns calculations within the Standard Model
framework, this will have the consequence that at the 
one loop level the independent
Lorentz structures of the invariant scattering amplitude will be of
only four types corresponding to initial and final axial and vector
"currents". Equivalently, one shall have, following the
definitions of ref.\cite{10}, a "$\gamma\gamma$", a "$ZZ$", a "$\gamma
Z$" and a "$Z\gamma$" structure, that will appear as the four
independent combinations of the elementary $\gamma, Z$ "currents"
defined as

 \bq v_{\mu f}^{(\gamma)}=
e_0Q_f\bar u_f\gamma_{\mu}v_f \eq

 \bq v_{\mu f}^{(Z)}=
{e_0\over 2c_0s_0}
\bar u_f\gamma_{\mu}(g^{(0)}_{Vf}-\gamma^5 g^{(0)}_{Af})v_f \eq

For instance, the "$\gamma Z$" structure will correspond in our
notations to the product of $v^{(\gamma)}_{\mu,l}v^{\mu(Z)}_{f}$,
while the "$Z\gamma$" structure will correspond to
$v^{(Z)}_{\mu,l}v^{\mu(\gamma)}_{f}$.\par

In this paper, we shall focuse our attention on three cases that we
consider to be realistic at a future $500~GeV$ electron-positron
collider, i.e. those of production of 
two final charged leptons ($A_{LR,l}$), of a
final $b\bar b$  couple ($A_{LR,b}$), and
of production of all possible light 
final quark couples ($A_{LR,5}$). This should cover all
the meaningful possibilities for two final light fermions
production.\par

The previous equations (1),(2) were strictly valid at Born level. To
make more rigorous statements, one has now to move to the one loop
expressions. This implies a redefinition of the various bare quantities
and also a consideration of the potentially dangerous QED radiation. For
what concerns the latter point, a rigorous treatment of $A_{LR,f}$
at NLC (on $Z$ resonance an exhaustive 
discussion is available \cite{11}), does
not yet exist to our knowledge (and is, in fact, under examination). We
shall assume that, as it happens in all other cases, a proper 
apparatus-dependent calculation 
allows to eliminate the unwanted difficulties
and we proceed from now on to the treatment of the purely "non QED"
component. In the latter one we shall leave aside, and consider it as a
separate and fixed component, the contribution to the considered
observables originated by standard strong interactions that, in the
conventional treatment, will be denoted as the "QCD" term $\simeq
\alpha_s(q^2)$. Our interest will be concentrated on the purely
electroweak components of the various $A_{LR,f}$, computed at one
loop. The color factor of all these quantities will consequently
continue to cancel exactly in the ratio, as it did at pure Born
level.\par

 To illustrate the philosophy and the main features of our approach
with the simplest example, we shall consider the modifications at one
loop of the "pure $Z$" Born exchange term $\simeq{1\over
(q^2-M^2_Z)^2}$ in the denominator of a general $A^{(0)}_{LR,f} \equiv
\sigma^{(0)}_{LR,f}/\sigma^{(0)}_{f}$. As it has been shown in full
detail in ref.\cite{10}, sect.2, in the discussion leading to
eq.(38), and as one can easily derive, the Born expression becomes at
one loop:

\bqa && \sigma^{(Z)}_{lf}(q^2)=N_f({4\pi q^2\over3}){[{3\Gamma_l\over
M_Z}][{3\Gamma_f\over N^{QCD}_f M_Z}]
\over(q^2-M^2_Z)^2+M^2_Z\Gamma^2_Z}[
1-2R^{(lf)}(q^2)\nonumber\\
&& -8s_lc_l\{{\tilde{v}_l\over1+\tilde{v}^2_l}V^{(lf)}_{\gamma
Z}(q^2)+{\tilde{v}_f
|Q_f|\over1+\tilde{v}^2_f}V^{(lf)}_{Z\gamma}(q^2)\}] \eqa

We want to stress now again the main features of this equation. As one
sees, the squared Fermi coupling $G^{(0)2}_{\mu}$ has been replaced by
the product of the two $Z$ widths $\Gamma_l\Gamma_f$, for which one is
supposed to take the experimental values measured on top of $Z$
resonance (in fact, for $f\neq l$, $\Gamma_f$ appears divided by the
quantity $N^{QCD}_f\simeq 3(1+\alpha_s 
(M^2_Z)/\pi)$, where also
$\alpha_s(M^2_Z)$ is supposed to be measured on top of $Z$ resonance).
As a consequence of this bargain, the one loop "form factors"
$R^{lf}(q^2)$, $V^{lf}(q^2)$ are \underline{subtracted} at $q^2=M^2_Z$.
More precisely, they will be given by integrations over the angular
variable of the following expressions:

\bq R^{(lf)}(q^2,\theta) \equiv \tilde{I}^{(lf)}_Z (q^{2}, \theta) -
\tilde{I}^{(lf)}_Z (M^2_Z, \theta) \eq

\bq V^{(lf)}_{\gamma Z}(q^2,\theta) \equiv \tilde{F}^{(lf)}
_{\gamma Z} (q^{2}, \theta) -
\tilde{F}^{(lf)}_{\gamma Z} (M^2_Z, \theta) \eq

\bq V^{(lf)}_{Z\gamma}(q^2,\theta) \equiv \tilde{F}^{(lf)}
_{Z\gamma} (q^{2}, \theta) -
\tilde{F}^{(lf)}_{Z\gamma} (M^2_Z, \theta) \eq

\noindent
where the "auxiliary" quantity $\tilde{I}_Z$ is defined as

\bq \tilde{I}^{(lf)}_Z(q^2,\theta) = 
{q^2\over q^2-M^2_Z}[\tilde{F}^{(lf)}_Z (q^2, \theta)-
\tilde{F}^{(lf)}_{Z} (M^2_Z, \theta)]  \eq

\noindent
and all the quantities denoted as $\tilde{F}^{(ij)}$ in the previous equations
are conventional, \underline{gauge invariant} combinations of
self-energies, vertices and boxes (defined following the conventions of
Degrassi and Sirlin \cite{12}) that belong to the previously defined
"$ZZ$", "$\gamma Z$" and "$Z\gamma$" Lorentz structures. To fix the
normalization, the \underline{self-energy} ($cos\theta$ independent)
component of $\tilde{F}^{ij}$ is the one appearing in the usual
definition of the transverse self-energies:

\bq A_{i}(q^{2}) \equiv A_{i}(0) + q^{2} F_{i} (q^{2}) \eq

 The generalization of the given example to the complete expressions of
the asymmetries will now be a trivial one. In the final
\underline{"$\gamma Z$" component}, for instance, new quantities
measured on $Z$ resonance will appear. One will be the longitudinal
polarization asymmetry itself, defined as ;

\bq A_{LR}(M^2_Z)={2\tilde{v}_{l}(M^2_Z)\over 1 
+ \tilde{v}^2_{l}(M^2_Z)}
\eq
\noindent
where $\tilde{v}_l(M^2_Z)=1-4s^2_l(M^2_Z)$ and $s^2_l(M^2_Z)$ is the
effective (leptonic) Weinberg-Salam angle measured at $M^2_Z$. Also,
the corresponding hadronic variables $\tilde{v}_f(M^2_Z)$ will enter,
whose \underline{exact} definition is provided by the so called
polarized forward-backward asymmetries originally \cite{13} defined as:

\bq A_{b,c} = {2\tilde{v}_{b,c}(M^2_Z)\over 1 + 
\tilde{v}^2_{b,c}(M^2_Z)} \eq

\noindent
(In practice, $\tilde{v}_f\simeq 1-4|Q_f|\tilde{s}^2_l(M^2_Z)$.)

One should also say at this point that, for what concerns the photon
contribution, the treatment is of strictly conventional type, with the
bare $\alpha^{(0)}$ replaced by the physical coupling computed at zero
momentum transfer $\alpha_{QED}\equiv \alpha(0)$ and the form factor

\bq \tilde{\Delta}^{(lf)}\alpha(q^2,\theta) \equiv 
\tilde{F}^{(lf)}_{\gamma} (0, \theta) 
- \tilde{F}^{(lf)}_{\gamma} (q^{2}, \theta)  \eq

\noindent
where $\tilde{F}_{\gamma}$ is a proper projection on the
"$\gamma\gamma$" structure of the usual photon self-energy with
corresponding vertices and boxes.\par

 After this long but, we hope, useful discussion we shall now write the
required one loop expressions of the electroweak component of the
considered asymmetries. Using the previous notations, we have that:

\bqa &&\sigma^{(1)}_{LR,f}=
N_f({4\pi q^2\over3})\{{[{3\Gamma_l\over
M_Z}][{3\Gamma_f\over N^{QCD}_f M_Z}]\over(q^2-M^2_Z)^2+M^2_Z\Gamma^2_Z}[
1-2R^{(lf)}(q^2)\nonumber\\
&& -{4s_lc_l\over \tilde{v}_l}V^{(lf)}_{\gamma
Z}(q^2)-{8s_lc_l\tilde{v_f}
|Q_f|\over1+\tilde{v}^2_f}V^{(lf)}_{Z\gamma}(q^2)]
+2\alpha(0)|Q_f|{q^2-M^2_Z\over
q^2((q^2-M^2_Z)^2+M^2_Z\Gamma^2_Z)}\nonumber\\
&& [{3\Gamma_l\over
M_Z}]^{1/2}[{3\Gamma_f\over N^{QCD}_f
M_Z}]^{1/2}{\tilde{v}_f\over(1+\tilde{v}^2_l)^{1/2}(1+
\tilde{v}^2_f)^{1/2}}[1+
\tilde{\Delta}^{(lf)}\alpha(q^2) -R^{(lf)}(q^2)\nonumber\\ 
&&-{4s_lc_l|Q_f|\over \tilde{v}_f}V^{(lf)}_{Z\gamma}(q^2)]\}
\eqa

\bqa  &&\sigma^{(1)}_f=N_f({4\pi q^2\over3})\{
Q^2_f
{\alpha^2(0)\over q^4}[1+2\tilde{\Delta}^{(lf)}\alpha(q^2)]
+{[{3\Gamma_l\over
M_Z}][{3\Gamma_f\over N^{QCD}_f M_Z}]\over(q^2-M^2_Z)^2+M^2_Z\Gamma^2_Z}[
1-2R^{(lf)}(q^2)\nonumber\\
&& -8s_lc_l\{{\tilde{v}_l\over1+\tilde{v}^2_l}V^{(lf)}_{\gamma
Z}(q^2)+{\tilde{v}_f
|Q_f|\over1+\tilde{v}^2_f}V^{(lf)}_{Z\gamma}(q^2)\}]
+2\alpha(0)|Q_f|{q^2-M^2_Z\over
q^2((q^2-M^2_Z)^2+M^2_Z\Gamma^2_Z)}\nonumber\\
&& [{3\Gamma_l\over
M_Z}]^{1/2}[{3\Gamma_f\over N^{QCD}_f
M_Z}]^{1/2}{\tilde{v}_l\tilde{v}_f\over(1+
\tilde{v}^2_l)^{1/2}(1+\tilde{v}^2_f)^{1/2}}[1+
\tilde{\Delta}^{(lf)}\alpha(q^2) -R^{(lf)}(q^2) \nonumber\\
&&  -{4s_lc_l\over\tilde{v}_l}V^{(lf)}_{\gamma
Z}(q^2)+
{|Q_f|\over\tilde{v}_f}V^{(lf)}_{Z\gamma}(q^2)]\} 
\eqa

Eqs.(14) and (15) conclude our introductory discussion. In the
forthcoming part of this section we shall consider in more detail the
various cases corresponding to the three chosen different final states.

\newpage
{\bf 2b. Discussion of different final states}
\vspace{0.5cm}

We begin with the (simplest) case of two final charged leptons. Since
$f=l$, only three independent form factors will remain ($V_{\gamma
Z}\equiv V_{Z\gamma}$). To maintain the notations of refs.\cite{9},
\cite{10} we
shall put no fermion index on them, so that they will be labelled as
$\tilde{\Delta}\alpha$, $R$ and $V$. Also, we shall use here the
quantity

\bq  \kappa\equiv {\alpha(0)\over [3\Gamma_l/M_Z]}
\eq
From eqs.(14).(15) one is then led to the desired expression. Here for
simplicity we shall write it in an "effective" way i.e. throwing away
terms that are numerically irrelevant and only retaining the meaningful
contributions. In this way we obtain the (relatively simple) formula:

\bqa &&A^{(1)}_{LR,l}(q^2)={q^2[\kappa(q^2-M^2_Z)+q^2]
\over \kappa^2(q^2-M^2_Z)^2+q^4}A_{LR}(M^2_Z)\times
\nonumber \\
&&\bigm\{1+[{\kappa(q^2-M^2_Z)
\over\kappa(q^2-M^2_Z)+q^2}-{2\kappa^2(q^2-M^2_Z)^2\over
\kappa^2(q^2-M^2_Z)^2+q^4}]
[\tilde{\Delta}\alpha(q^2)+R(q^2)]
-{4c_ls_l\over \tilde{v}_l}V(q^2) \bigm\} \eqa

A few comments are, at this point, appropriate. First of all, one sees
that numerically the value of eq.(17) (more precisely, of its leading
term, the first one in the r.h.s. of the equation) decreases when $q^2$
becomes larger than $M^2_Z$, pointing to an asymptotic value of about
${1\over 2}A_{LR}(M^2_Z)\simeq 0.07$. The one loop modifications to the
leading term contain two quantities, the combination
$[\tilde{\Delta}\alpha + R]$ and the "$\gamma Z$" term $V$. The fact
that the sum $[\tilde{\Delta}\alpha + R]$ appears in eq.(17) is not
accidental: it will be a general feature for the new physics effects in 
any ratio of cross sections. We shall return on this point in the next
section. The point that deserves attention is the fact that the
coefficient of $V$ is relatively enhanced with respect to the
coefficient of $[\tilde{\Delta}\alpha + R]$
by the factor ${1\over \tilde{v}_l}$, which
makes it one order of magnitude larger. Note that this fact 
comes from the (accidental) smallness of the quantity
$\tilde{v}_l(M^2_Z)$ and is generated by the contribution to the
"$\gamma Z$" structure in the "pure $Z$" exchange component of
$\sigma_{LR,l}$ \underline{at one loop}, that is
strongly reminiscent of the situation met on top of $Z$ resonance. As a
consequence of this, one expects a relative enhancement of the
virtual effects for those models of new physics where the contribution
to $V(q^2)$ is not accidentally depressed. We shall provide one specific
example in section 3.\par
The next case that we shall consider is that of a final $b\bar b$ 
couple. From the relevant expressions, 
making the same numerical approximations 
as in the previous case i.e. only 
retaining the dominant contributions to the various coefficients, 
we obtain in this case:

\bq   A^{(1)}_{LR,b}(q^2)=\bar A_{LR,b} (q^2)[1+a_b
(q^2)[\tilde{\Delta}^{lb}\alpha(q^2)
+R^{lb}(q^2)]+b_b (q^2)V^{lf}_{\gamma Z}(q^2)+
c_b (q^2)V^{lf}_{Z\gamma}(q^2)]
\eq
\noindent
where
\bq
\bar A_{LR,b} (q^2)= {C_{LR,b}(q^2)\over C_b(q^2)}
\eq
\bq
a_b (q^2)=C^{\gamma Z}_{LR,b}-2C^{\gamma\gamma}_{b}-C^{\gamma Z}_{b}
=-2C^{ZZ}_{LR,b}-C^{\gamma Z}_{LR,b}+
2C^{ZZ}_{b}+C^{\gamma Z}_{b}
\eq
\bq
b_b (q^2)=-{4s_lc_l\over \tilde{v}_l}[C^{ZZ}_{LR,b}-C^{\gamma Z}_{b}]
+{8s_lc_l\tilde{v}_l\over1+\tilde{v}^2_l}C^{ZZ}_{b}
\eq
\bq
c_b (q^2)=-{4s_lc_l\over 3\tilde{v}_b}[C^{\gamma Z}_{LR,b}
-C^{\gamma Z}_{b}]
+{8s_lc_l\tilde{v}_b\over3(1+\tilde{v}^2_b)}[C^{ZZ}_{b}-C^{ZZ}_{LR,b}]
\eq
\noindent

and
\bq
C_{LR,b}=N^{ZZ}_{LR,b}+N^{\gamma Z}_{LR,b}
\eq
\bq
N^{ZZ}_{LR,b}={18\tilde{v}_l\over1+\tilde{v}^2_l}({3\Gamma_l/M_Z\over
\alpha})({3\Gamma_b/N^{QCD}_b M_Z\over\alpha})({q^4
\over(q^2-M^2_Z)^2+M^2_Z\Gamma^2_Z})
\eq
\bq
N^{\gamma
Z}_{LR,b}={6\tilde{v}_b\over(1+\tilde{v}^2_l)^{1/2}
(1+\tilde{v}^2_b)^{1/2}}
({{3\Gamma_l/M_Z}^{1/2}\over
\alpha})([3\Gamma_b/N^{QCD}_b M_Z]^{1/2})({q^2(q^2-M^2_Z)
\over(q^2-M^2_Z)^2+M^2_Z\Gamma^2_Z})
\eq
\bq
C^{ZZ}_{LR,b}= N^{ZZ}_{LR,b}/C_{LR,b}
\ \ \ \ \
C^{\gamma Z}_{LR,b}=N^{\gamma Z}_{LR,b}/C_{LR,b}
\ \ \ \ \
C_{b}=1+N^{ZZ}_{b}+N^{\gamma Z}_{b}
\eq
\bq
N^{ZZ}_{b}=9({3\Gamma_l/M_Z\over
\alpha})({3\Gamma_b/N^{QCD}_b M_Z\over\alpha})({q^4
\over(q^2-M^2_Z)^2+M^2_Z\Gamma^2_Z})
\eq
\bq
N^{\gamma
Z}_{b}={6\tilde{v}_l\tilde{v}_b\over(1+\tilde{v}^2_l)^{1/2}
(1+\tilde{v}^2_b)^{1/2}}
({{3\Gamma_l/M_Z}^{1/2}\over
\alpha})([3\Gamma_b/N^{QCD}_b M_Z]^{1/2})({q^2(q^2-M^2_Z)
\over(q^2-M^2_Z)^2+M^2_Z\Gamma^2_Z})
\eq
\bq
C^{\gamma\gamma}_{b}=1/C_{b}
\ \ \ \ \
C^{ZZ}_{b}= N^{ZZ}_{b}/C_{b}
\ \ \ \ \
C^{\gamma Z}_{b}=N^{\gamma Z}_{b}/C_{b}
\eq

Comparing eq.(18) with the previous eq.(17), we notice the following
facts. The numerical value of the leading term becomes asymptotically,
as one easily sees, much larger than that of the corresponding leptonic
quantity. For large $q^2$, using the experimental inputs for the
various widths and asymmetries, it approaches a value of approximately
$0.65$. In the one loop corrections, the largely dominant coefficient
is that of $V_{\gamma Z}$, approximately one order of magnitude larger
than that of $[\tilde{\Delta}\alpha + R]$, again as a consequence of
the $\simeq{1\over \tilde{v}_l}$ factor. Note that the enhanced
coefficient, i.e. that of $V_{\gamma Z}$, comes from the
\underline{pure Z} exchange contribution to $\sigma_{LR,b}$, where
$\tilde{v}_l$ appears at Born level. This feature, that is not valid
for the crossed term $V_{Z\gamma}$ where $\tilde{v}_l$ is replaced by
the much bigger $\tilde{v}_b$,
is the same that has been already met in the case of the leptonic
asymmetry.\par
The final case to be considered is that of the full longitudinal
asymmetry for production of the five light quarks
$A_{LR,5}=\sigma_{LR,5}/\sigma_5$. To derive its expression is
straightforward once the prescriptions of our approach have been made
clear. In pratice, the only new experimental quantities that will enter
in the theoretical formulae will be the $c$ asymmetry on $Z$ resonance
and the overall Z width $\Gamma_5$. The various relevant expressions
have all been given and computed in Ref.\cite{10}, 
where it has also been
shown that the related experimental error would not produce any
consequence in the theoretical formulae for unpolarized quantities 
that contain them as an input.
We shall return on this point at the end of the paper. For the moment
we write the final expression for the asymmetry introducing a
separation of the new physics effects that will be useful for our next
analysis. More precisely, we define systematically, for any model of
new physics and final state $f$:

\bq \tilde{\Delta}^{(lf)}\alpha(q^2) = 
\tilde{\Delta}\alpha(q^2) +
\delta\tilde{\Delta}^{(lf)}\alpha(q^2) \eq

\bq R^{(lf)}(q^2) = R(q^2) + \delta R^{(lf)}(q^2) \eq

\bq V^{(lf)}_{\gamma Z}(q^2) = V(q^2) +
\delta V^{(lf)}_{\gamma Z}(q^2) \eq

\bq V^{(lf)}_{Z\gamma}(q^2) = V(q^2) +
\delta V^{(lf)}_{Z\gamma}(q^2) \eq

\noindent
where the first bracket contains the "universal" (without index)
effects, i.e. those that would be exactly the same for final leptons or
quarks.\par

Using the previous definitions, it is relatively easy to derive the
expression of $A_{LR,5}$ for models of new physics that are of
\underline{universal type}. Working in the usual spirit of only
retaining the important contributions we would obtain the following
formula

\bq  A^{(1)}_{LR,5}(q^2)= \bar A_{LR,5}(q^2)\{1+a_5
(q^2)[\tilde{\Delta}\alpha(q^2)
+R(q^2)]+[b_5 (q^2)+c_5 (q^2)]V(q^2)\}
\eq
\noindent
where

\bq
\bar A_{LR,5} (q^2)= {C_{LR,5}(q^2)\over C_5(q^2)}
\eq
\bq
a_5 (q^2)=C^{\gamma Z}_{LR,5}-2C^{\gamma\gamma}_{5}-C^{\gamma Z}_{5}
=-2C^{ZZ}_{LR,5}-C^{\gamma Z}_{LR,5}+
2C^{ZZ}_{5}+C^{\gamma Z}_{5}
\eq

\bq
b_5 (q^2)+c_5 (q^2)=-4s_lc_l\{[p_{LR,5}C^{ZZ}_{LR,5}
+p'_{LR,5}C^{\gamma Z}_{LR,5}]
-[p_5 C^{ZZ}_{5}+p'_5 C^{\gamma Z}_{LR,5}]\}
\eq
\noindent
and
\bq
C_{LR,5}=N^{ZZ}_{LR,5}+N^{\gamma Z}_{LR,5}
\eq
\bq
N^{ZZ}_{LR,5}=({2\tilde{v}_l\over1+\tilde{v}^2_l})
{[{3\Gamma_l\over M_Z}]
[{3\Gamma_5\over M_Z}]\over(q^2-M^2_Z)^2+M^2_Z\Gamma^2_Z}
\eq
\bq
N^{\gamma
Z}_{LR,5}={2\alpha\over3(1+\tilde{v}^2_l)^{1/2}}[{3\Gamma_l\over
M_Z}]^{1/2}
\Sigma_5{ q^2(q^2-M^2_Z)\over(q^2-M^2_Z)^2+M^2_Z\Gamma^2_Z}
\eq

\bq
C^{ZZ}_{LR,5}= N^{ZZ}_{LR,5}/C_{LR,5}
\ \ \ \ \
C^{\gamma Z}_{LR,5}=N^{\gamma Z}_{LR,5}/C_{LR,5}
\ \ \ \ \
C_{5}=1+N^{ZZ}_{5}+N^{\gamma Z}_{5}
\eq

\bq
N^{ZZ}_{5}=(\frac{9}{33})[{3\Gamma_l/ M_Z\over\alpha}]
[{3\Gamma_5/ M_Z\over\alpha}]{q^4\over(q^2-M^2_Z)^2+M^2_Z\Gamma^2_Z}
\eq
\bq
N^{\gamma Z}_{5}=(\frac{2}{11}){\tilde{v}_l\over(1+\tilde{v}^2_l)^{1/2}}
{[3\Gamma_l/ M_Z]^{1/2}\over\alpha}
\Sigma_5
{q^2(q^2-M^2_Z)\over(q^2-M^2_Z)^2+M^2_Z\Gamma^2_Z}
\eq
\bq  \Sigma_5=
\sum_q{3|Q_q|\tilde{v}_q\over(1+\tilde{v}^2_q)^{1/2}}({3N_q\Gamma_q
\over M_Z})^{1/2}
\eq

\bq
C^{\gamma\gamma}_{5}=1/C_{5}
\ \ \ \ \
C^{ZZ}_{5}= N^{ZZ}_{5}/C_{5}
\ \ \ \ \
C^{\gamma Z}_{5}=N^{\gamma Z}_{5}/C_{5}
\eq

\bq  p_5= {\tilde{v}_l\over1+\tilde{v}^2_l}+
\sum_q({\tilde{v}_q\over1+\tilde{v}^2_q}){|Q_q|\Gamma_q\over\Gamma_5}
\ \ \ \ \ \ p'_5=  {1\over
\tilde{v}_l}+ p'_{LR,5}
\eq
\bq
p'_{LR,5}=\sum_q({3|Q_q|^2\over\Sigma_5(1+\tilde{v}^2_q)^{1/2}})
({3N_q\Gamma_q\over M_Z})^{1/2}
\ \ \ \ \ \ 
p_{LR,5}= {1\over\tilde{v}_l}+\sum_q({\tilde{v}_q\over1
+\tilde{v}^2_q})
{2|Q_q|\Gamma_q\over3\Gamma_5}
\eq

The coefficient $\bar A_{LR,5}(q^2)$, 
in this particular notation, contains both
the leading ("effective" Born) terms and the one-loop corrections
$\delta \bar A^{SM}_{LR,5}$ of the pure SM. 
The latter ones will not be, in
general, of universal type, since they involve vertices and boxes.
Neglecting their numerical value for a first estimate of the leading
term gives us the expected large $q^2$ value of the asymmetry, that is
approximately $\bar A_{LR,5}(q^2)\simeq 0.50$. 
For what concerns the remaining
coefficients, one easily sees that, once again, that of
$[\tilde{\Delta}\alpha + R]$ is more than one order of magnitude
smaller than that of $V$. The latter one, in turn, comes mostly from
the \underline{universal} component of $V_{\gamma Z}$ reproducing the
situation that we have already met in the two previous examples.\par

 This recurrent feature of "$V_{\gamma Z}$" dominance of the one loop
effects of new physics survives, in the last considered asymmetry, even
in the most general case of non universal type of effects, as one can
see if one writes the full expression that generalizes eq.(34) to this
case. This can be done in a straightforward way, and leads to the
rather lengthy expression that we write here for completeness:

\bqa && A^{(1)}_{LR,5}(q^2)= \bar A_{LR,5}(q^2)\{1+a_5
(q^2)[\tilde{\Delta}\alpha(q^2)
+R(q^2)]+[b_5 (q^2)+c_5 (q^2)]V(q^2)\nonumber\\
&&+\sum_q\delta\tilde{\Delta}^{(lq)}\alpha(q^2)
[(C^{\gamma Z}_{LR,5}(q^2)-C^{\gamma Z}_{5}(q^2)){3|Q_q|\tilde{v}_q
\over\Sigma_5(1+\tilde{v}^2_q)^{1/2}}
({3N_q\Gamma_q\over M_Z})^{1/2}
-C^{\gamma\gamma}_5(q^2)\frac{18}{11}|Q_q|^2]\nonumber\\
&&+\sum_q\delta
R^{(lq)}(q^2)[2(C^{ZZ}_{5}(q^2)-C^{ZZ}_{LR,5}
(q^2)){\Gamma_q\over \Gamma_5}
+(C^{\gamma Z}_{5}(q^2)-C^{\gamma Z}_{LR,5}(q^2)){3|Q_q|\tilde{v}_q
\over\Sigma_5(1+\tilde{v}^2_q)^{1/2}}
({3N_q\Gamma_q\over M_Z})^{1/2}]\nonumber\\
&&+\sum_q\delta V^{(lq)}_{\gamma Z}(q^2)[
({8s_lc_l\tilde{v}_l\over1+\tilde{v}^2_l})
C^{ZZ}_{5}(q^2){\Gamma_q\over \Gamma_5}-{4s_lc_l\over \tilde{v}_l}
(C^{ZZ}_{LR,5}(q^2)2{\Gamma_q\over \Gamma_5}
-C^{\gamma Z}_{5}{3|Q_q|\tilde{v}_q
\over\Sigma_5(1+\tilde{v}^2_q)^{1/2})}
({3N_q\Gamma_q
\over M_Z})^{1/2})]\nonumber\\
&&+\sum_q\delta V^{(lq)}_{Z\gamma}(q^2)][
{|Q_q|4s_lc_l\over\tilde{v}_q}(C^{\gamma Z}_{5}(q^2)
-C^{\gamma Z}_{LR,5}(q^2))
{3|Q_q|\tilde{v}_q
\over\Sigma_5(1+\tilde{v}^2_q)^{1/2})}
({3N_q\Gamma_q
\over M_Z})^{1/2}\nonumber\\
&&+({8s_lc_l|Q_q|\tilde{v}_q\over1+\tilde{v}^2_q})
(C^{ZZ}_{5}(q^2)-C^{ZZ}_{LR,5}(q^2)){\Gamma_q\over \Gamma_5}]\}
\eqa
\noindent
and, this time, the one-loop corrections contain both the SM and the
new physics effects. Note that the $[\tilde{\Delta}\alpha(q^2)
+R(q^2)]$ combination only appears for the universal term. 
Leaving aside a more quantitative discussion in
this non-universal case, we only remark that, 
as we said previously, the weight of
both the universal and the non-universal
$V^{lf}_{\gamma Z}$ components remain essentially 
enhanced by the typical
$1/\tilde{v}_l$ effect, that remains in conclusion the relevant
feature of all the considered longitudinal polarization asymmetries.\par

This characteristic feature should be compared now with those of other
specific unpolarized observables. We have done this for the following
relevant quantities, exploiting their theoretical expressions in our
approach that can be found in refs.\cite{9},\cite{10}:

I) $A_{FB,\mu}$, the muon forward-backward asymmetry. Here the size of
the coefficient of the sum $[\tilde{\Delta}\alpha + R]$ that still
appears as a unique block is approximately three times bigger than that
of $V$.

II) $\sigma_{\mu}$, the muon cross-section. Here the dominant effect is
by far (one order of magnitude) concentrated in the correction
$\tilde{\Delta}\alpha$ (that now is no more related to $R$ as in the
previous asymmetries).

III) $\sigma_5$, the five light quark cross section. For the case of
\underline{universal} effects, the coefficients of all the three form
factors $\tilde{\Delta}\alpha$, $R$ and $V$ are now roughly equal (this
remains qualitatively true for general non universal effects).

IV) $\sigma_b$, the $b\bar b$ cross section. Here, the leading
coefficients of nearly equal size are those of $R$ and $V$.\par

This short analysis shows that, indeed, longitudinal polarization
asymmetries are much more sensitive to one specific one-loop effect
$\simeq V_{\gamma Z}$ and therefore to all those models that contribute
this quantity in a sensible way. One the contrary, unpolarized leptonic
observables are a better place for looking at effects generated by
either the combination $[\tilde{\Delta}\alpha + R]$ (e.g. $A_{FB,\mu}$) or
by the separate quantity  $\tilde{\Delta}\alpha$ (e.g. $\sigma_{\mu}$). In
unpolarized hadronic quantities, the three form factors 
$\tilde{\Delta}\alpha$, $R$ and $V$ all appear with coefficients of
similar size.\par
We still have to discuss three specific points. The first one is that,
as previously stressed, it is the combination 
$[\tilde{\Delta}\alpha + R]$ that appears systematically in ratios of
cross sections. This can be understood from the general ($\gamma$,
$Z$) structure if we write the general
expression of any cross section in the following way

\bq    
\sigma^{lf}_1\equiv c^{\gamma\gamma}_1(1+2\tilde{\Delta}^{lf}\alpha)
+c^{ZZ}_1(1-2R^{lf})+c^{\gamma Z}_1(1+\Delta^{lf}\alpha-R^{lf}) +\ \  V~
terms
\eq

\noindent
defining $c_1\equiv  c^{\gamma\gamma}_1+c^{ZZ}_1+c^{\gamma Z}_1$, one
can write
\bq 
\sigma^{lf}_1\equiv c_1[1+\tilde{\Delta}^{lf}\alpha
-R^{lf}+{ c^{\gamma\gamma}_1
-c^{ZZ}_1\over c_1}(\tilde{\Delta}^{lf}\alpha)+R^{lf})] +\ \   V~terms
\eq
 Keeping only first order terms in
$\tilde{\Delta}^{lf}\alpha$, $R^{lf}$ and $V$,
the ratio of two such cross sections $\sigma^{lf}_1$ and
$\sigma^{lf}_2$, is given by:

\bq {\sigma^{lf}_1\over \sigma^{lf}_2} = {c_1\over c_2}[1+
({ c^{\gamma\gamma}_1-c^{ZZ}_1\over c_1}-{
c^{\gamma\gamma}_2-c^{ZZ}_2\over c_2})(\tilde{\Delta}^{lf}\alpha
+R^{lf})] +\ \   V~terms
\eq
\noindent
in which only the combination $[\tilde{\Delta}^{lf}\alpha
+R^{lf}]$ appears. Note that this property in general disappears 
if one considers
ratios of sums of different flavors 
$(\sum_f\sigma^{lf}_1)/(\sum_f\sigma^{lf}_2)$, as we have seen in the
case of $A^{(1)}_{LR,5}$, eq.(48).\par

The second point is the statement that, in order to exploit the
properties of $A_{LR,f}$, the contribution of the model of new physics
to $V$ must not be accidently depressed with respect to that of
$[\tilde{\Delta}\alpha + R]$. Although we cannot prove this fact in
general, we shall provide now in the next section a specific example of
a model where this is actually the case, and for which the role of
$A_{LR,f}$ will consequently be very useful.\par

The final point is that of whether the bargain introduced in our
approach by the replacement of $G_{\mu}$ with $Z$-peak quantities does
not generate a dangerous theoretical input error (in the case of
unpolarized observables, this was shown not to be the case for future
$e^+e^-$ colliders at their realistically expected accuracy in
refs.\cite{9},\cite{10}. Let us start with the leptonic asymmetry
eq.(17). In our approach, its new theoretical expression at the
"effective" Born level is the first member on the r.h.s. of eq.(17).
This contains the $Z$ leptonic width $\Gamma_l$ and $A_{LR}$ measured
on top of $Z$ resonance. With the available errors on these quantities,
one computes a theoretical error in eq.(17) of approximately $0.0018$
which is mostly coming from $A_{LR}$. Assuming an (optimistic)
experimental error on $A_{LR}$ at a $500~GeV$ NLC \cite{NLC} of $0.007$
(purely statistical) one sees that the induced theoretical error is
completely negligible. This statement will also 
be made more accurate by future
improvements on the measurement of $A_{LR}$ at SLD
\cite{7},\cite{SLDLecce}.\par
In the case of eq.(18), one easily sees that the major source of error
in the expression of the "effective" Born terms comes from the quantity
${\tilde{v}_b\over \sqrt{1+\tilde{v}^2_b}}$. To compute this error, we
have used the definition eq.(12), from which we obtain:

\bq \delta
\tilde{v}_b={(1+\tilde{v}^2_b)^2\over2(1-\tilde{v}^2_b)}\delta A_b
\eq
\noindent 
Using the experimental LEP+SLD value \cite{7}, \cite{lastres}

\bq   A_b= 0.867\pm0.022  
\eq
\noindent
we derive $\delta\tilde{v}_b\simeq 0.04$. A standard calculation then
gives for the theoretical input error:

\bq  \delta^{(th)} \bar A_{LR,b}(q^2)\simeq 0.02
\eq

Note that this numerical result is directly proportional to the
experimental error on $A_b$, and will be correspondingly reduced by
future improved measurements of this quantity. This final error should
be compared to the expected experimental precision on $A_{LR,b}$.
Although a detailed study does not exist yet to our knowledge, we can
reasonably foresee a picture for $b\bar b$ detection similar to that
found for previous LEP2 studies \cite{LEP2}, 
that would lead to an overall
error of at least a few percent, sufficiently larger than our
theoretical input error.\par

To conclude, we have considered the case of $A_{LR,5}$. This case can
be treated in a reasonably simple way since in the theoretical
expression of the leading term $\bar A_{LR,5}(q^2)$ 
the only relevant theoretical
uncertainty affects the "$\gamma Z$" component of the numerator (for
the denominator, a previous discussion given in ref.\cite{10} has shown
that the main error is coming from $\Gamma_h$, the $Z$ hadronic width
measured on $Z$ peak, and is completely negligible i.e. much smaller
than a relative one percent). The $\gamma Z$ component contains the $Z$
peak quantities $\Gamma_{u,d,s,c,b}$ and the related quantities 
$A_{u,d,s,c,b}$ defined by a generalization of eq.(12). In fact, no
experimental information is available on the ($u,d,s$) variables. A
reasonable attitude seems to us to be that of assuming a universality
property, i.e.

\bq   \Gamma_u=\Gamma_c \ \ \ \ \ \Gamma_d=\Gamma_s=\Gamma_b(m_t=0)
\eq
\noindent
and to derive $\Gamma_b(m_t=0)$ from its knowns experimental value
where the theoretical top quark contribution has been subtracted.
Analogously, we shall assume that $A_u=A_c$ and  $A_d=A_s=A_b(m_t=0)$
and for the latter quantity we have again subtracted the known (and
relatively small) top quark contribution. With these assumptions, one
easily sees that the major theoretical error is coming from that of
$\tilde{v}_c$ and $\tilde{v}_b$ (the induced error by the widths is
much smaller). Using the experimental SLD results for $A_{b,c}$ then
leads to an error of $A_{LR,5}$:

\bq  \delta^{th} \bar A_{LR,5}(q^2)\simeq 0.02
\eq

This is not a very comfortable result, since one would expect an
experimental error on $A_{LR,5}$ at NLC not far from the purely
statistical one, that is around one percent. In order to reduce the
theoretical error of our input to such values, an extra effort from SLD
that reduces to the one percent level the error on $A_b$ and to
the three percent level that on $A_c$ would be requested. Such a
desirable goal seems to be reachable in future SLD measurements
\cite{7}. In the rest of this
paper, we shall illustrate as an application the consequences of having
been able to reduce the overall error on $A_{LR,5}$ to the one
percent level. This will be done immediately in the next Section 3.\par

\section{A model with anomalous gauge couplings}

To illustrate the previous considerations with a concrete example, we
shall now consider the case of a model where anomalous gauge couplings
(AGC) are present. To be more precise, we shall discuss the
consequences of our approach for the study of a model proposed by
Hagiwara et al \cite{Hag}, to whose paper we defer for a full discussion
of various theoretical aspects. Briefly, the model assumes that physics
below a scale $\Lambda$ of supposed order $1~TeV$ can be described by
an "effective" Lagrangian obtained by adding to the conventional SM
component an extra $SU(2)\times U(1)$ invariant, C and CP conserving,
dimension six piece. The latter contains, a priori, eleven parameters
of which four enter at the one loop level for production of two final
\underline{massless} fermions from electron-positron annihilation. In
the notation of ref.\cite{Hag} these are called $f_{DW}$, $f_{DB}$,
$f_{BW}$ and $f_{\Phi,1}$. In a conventional treatment that does not
use our $Z$-peak representation they would all contribute this process
at one loop. The treatment of this model in our approach turns out to
be particularly convenient. As it has already been shown in
ref.\cite{9}, the number of effective parameters that appear in the
subtracted form factors is reduced to two ($f_{DW}$ and $f_{DB}$) since
$f_{BW}$ and $f_{\Phi,1}$ are fully reabsorbed in the used input
parameters $\Gamma_l$ and $s^2_l(M^2_Z)$. Another welcome feature of
this model is that its effects for massless fermions are of universal
type, so that the \underline{same} two parameters will enter both
leptonic and quark observables. This allows to determine informations
on bounds on these parameters in a greatly simplified way, using
several measurements of different experimental quantities. This was
done in a very recent paper \cite{clean} where the bounds that would be
obtained from negative searches both at LEP2 and at NLC without
polarization were derived. In
Fig.1 the results of that investigation are presented showing the
region within which the two parameters ($f_{DW}$, $f_{DB}$) would be
constrained by negative searches in the unpolarized case. Numerically,
we would find in this case:

\bq  \Delta  f_{DB}=\pm 0.16
\eq

\bq  \Delta f_{DW}=\pm 0.025
\eq

 In practice, the determination of such bounds in Fig.1 is mostly
provided by two quantities i.e. the muon cross section and the five
light hadron production cross section $\sigma_5$ (the 
forward-backward asymmetry $A_{FB,\mu}$ plays a
negligible role because of a weaker sensitivity). Their expression in
the considered model are provided in ref.\cite{clean} and are fixed by
the (AGC) content of the three form factors $\tilde{\Delta}\alpha$, $R$
and $V$, that read respectively:

\bq \tilde{\Delta}^{(AGC)}\alpha(q^2)= -q^2({2e^2\over\Lambda^2})
(f^r_{DW}+f^r_{DB}) \eq

\bq  R^{(AGC)}(q^2)= (q^2-M^2_Z)({2e^2\over s^2_lc^2_l\Lambda^2})
(f^r_{DW}c^4_l+f^r_{DB}s^4_l) \eq

\bq V^{(AGC)}(q^2) = (q^2-M^2_Z)({2e^2\over s_lc_l\Lambda^2})
(f^r_{DW}c^2_l-f^r_{DB}s^2_l)  \eq

We have now added to the previous unpolarized information that
derivable from longitudinal polarization asymmetries. To avoid problems
related to $b$ quark identification and to stick more rigorously to the
massless quark configuration, we have only considered the leptonic and
the full light hadronic asymmetry (where the weight of the $b$
contribution is sufficiently depressed). For the latter ones we have
assumed, following our previous discussion (and an optimistic attitude),
an experimental error $\delta A_{LR,l}=\pm 0.007$
and $\delta A_{LR,5}=\pm 0.01$. This is based on an integrated
luminosity of $20~fb^{-1}$ leading at $\sqrt{q^2}=500~GeV$ to about
$5\times10^{4}$ hadronic events and $1.7\times10^{4}$ (muon + tau
events). To give a hint of how the "$V$ enhancement"
mechanism works, we write the two corresponding theoretical expressions
in the chosen configuration $\sqrt{q^2}=500~GeV$, that numerically
read:\\
from $A_{LR,5}$
\bq  |{32\pi\alpha(0)M^2_Z\over\Lambda^2}[-53.36f_{DW} +
14.43f_{DB}]|=|{\delta A_{LR,5}\over  A_{LR,5}}|\gsim0.02
\eq 
\noindent
and from $A_{LR,l}$

\bq  |{32\pi\alpha(0)M^2_Z\over\Lambda^2}[-342.65f_{DW} + 92.55f_{DB}]|=
|{\delta A_{LR,l}\over  A_{LR,l}}|\gsim0.1
\eq
\noindent
(For $\Lambda=1~TeV$,
the coefficient 
$32\pi\alpha(0)M^2_Z\over\Lambda^2$ is equal to 0.0061).\par
In eqs.(62) and (63)
the last numbers on the r.h.s. represent the visibility threshold for
the effect. Note that both equations involve the same type of
combination of $f_{DW}$ and $f_{DB}$ couplings.  With the expected
accuracies, eq.(63) due to $A_{LR,l}$ is slightly more stringent than
eq.(62) due to $A_{LR,5}$. This is fortunate because of the
uncertainty on the final accuracy that will be reachable
on $A_{LR,5}$. In the following
numerical analysis we shall combine quadratically the informations
coming from these two constraints and this reduces somewhat the
importance of $A_{LR,5}$.

These expressions should be compared with those provided by the
unpolarized observables. Taking for simplicity the two most sensitive
quantities i.e. the muon and the hadron cross sections, the
corresponding equations would be:\\
from $\sigma_{\mu}$
\bq  |{32\pi\alpha(0)M^2_Z\over\Lambda^2}[-22.02f_{DW}
-13.07f_{DB}]|=|{\delta\sigma_{\mu}\over \sigma_{\mu}}|\gsim0.01
\eq
\noindent
and from $\sigma_{5}$

\bq  |{32\pi\alpha(0)M^2_Z\over\Lambda^2}[-49.53f_{DW}
-5.45f_{DB}]|=|{\delta\sigma_{5}\over \sigma_{5}}|\gsim0.005
\eq

Comparing eqs.(64)(65) with eqs.(62),(63) one actually sees that the
combination of $f_{DW}$, $f_{DB}$ that appear in the two sets are
almost orthogonal. This corresponds indeed, as we discussed in Section
2, to the fact that different form factors are selected in the two
cases.\par

From a practical point of view, the additional improvements for future
negative bounds derivable from the addition of the two extra
asymmetries is shown in Fig.1 . As one sees, the final limits would be:

\bq  \Delta  f_{DB}=\pm 0.08
\eq

\bq  \Delta f_{DW}=\pm 0.014
\eq

In other words, the additional constraint provided by longitudinal
polarization would lead, in this example, to an improvement in the
bounds equal to, roughly, a factor of two.

\section{Conclusions}

We have shown in this paper that longitudinal polarization asymmetries
of electron-positron annihilation into pairs of light
fermion-antifermion at energies 
larger than $M_Z$ exhibit interesting theoretical features that might
be useful for detection of a certain type of virtual effects of new
physics at one loop, and that are due to a special enhancement of the
subtracted $V$ form factor. This feature is analogous to that found on
top of $Z$ resonance, showing that $A_{LR}$ continues to be a relevant
observable even far from that privilegded kinematical configuration.\par

We have presented in this paper only one concrete example of how
this enhancement mechanism works, for the special case of one model of
universal type. Other similar cases could be examined. For instance,
general models of technicolour type (already qualitatively considered
in ref.\cite{9}) would probably benefit from a more detailed numerical
calculation. This will be done in a separate work. Also, the more
complicated case of models of non universal type would deserve
consideration. An interesting case would be that of general
supersymmetric models. Here the virtual effects are usually depressed
on $Z$ resonance. Away from $Z$ resonance, there might be, though,
unconventional effects of non universal type (we have in mind e.g.
boxes, that are kinematically depressed on $Z$ peak but resuscitate
when $(q^2-M^2_Z)$ is sufficiently large). These would enter in our
subtracted form factors at large energies since they would
\underline{not} be reabsorbed, by definition, in the $Z$ peak
observables that are the new inputs of our procedure. The study of this
possibility is by now in progress.

\vspace{0.5cm}
\leftline{\Large \bf Acknowledgments}
One of us (C.V.) wishes to thank the Department of Physique
Math\'ematique et Th\'eorique of Montpellier, where this work was
written, for the warm and friendly hospitality.

\newpage
\begin{center}

{\large \bf Figure caption}
\end{center}
\vspace{0.5cm}
{\bf Fig.1} Constraints on AGC couplings from $e^+e^-\to f\bar f$ 
processes at 500 GeV. 
Without polarization: $\sigma_{\mu}$ ($\Box$)~,~~ $\sigma_5$
($+$)~,~~$\sigma_b$~~($\times$).
With polarization: ~$A_{LR}$~~($\Diamond$) for which the band is
obtained by combining quadratically the informations 
coming from $A_{LR,5}$ and $A_{LR,l}$.

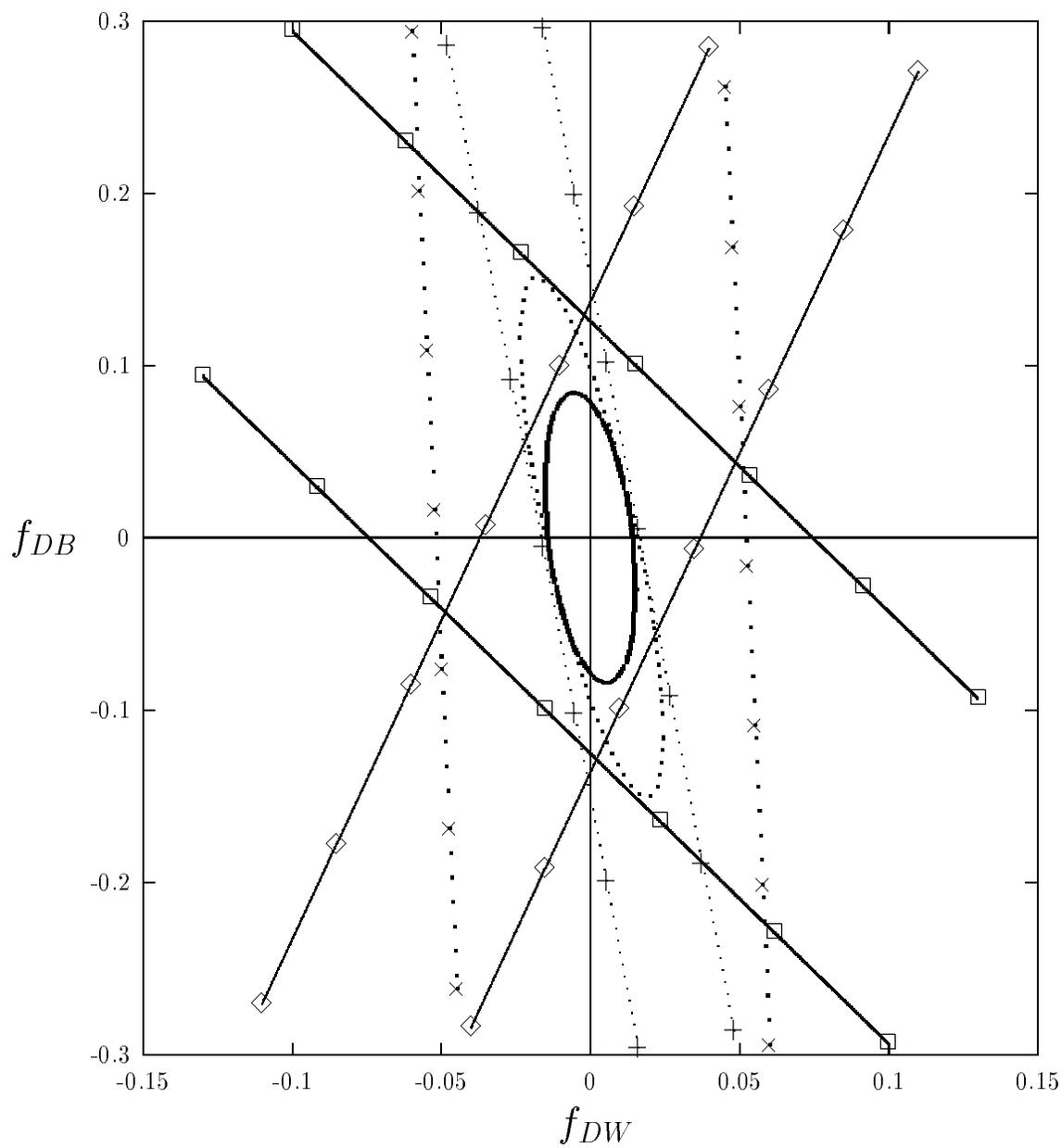
\end{document}